\documentclass[journal=nalefd,manuscript=preprint]{achemso} %,layout=twocolumn

\usepackage[version=3]{mhchem} % Formula subscripts using \ce{}
\usepackage{units}
\usepackage[section]{placeins}

\author{Marc Dr\"ogeler}
\affiliation[RWTH Aachen University]{2nd Institute of Physics and JARA-FIT, RWTH Aachen University, 52074 Aachen, Germany, EU}
\author{Frank Volmer}
\affiliation[RWTH Aachen University]{2nd Institute of Physics and JARA-FIT, RWTH Aachen University, 52074 Aachen, Germany, EU}
\author{Maik Wolter}
\affiliation[RWTH Aachen University]{2nd Institute of Physics and JARA-FIT, RWTH Aachen University, 52074 Aachen, Germany, EU}
\author{Bernat Terr\'{e}s}
\affiliation[RWTH Aachen University]{2nd Institute of Physics and JARA-FIT, RWTH Aachen University, 52074 Aachen, Germany, EU}
\alsoaffiliation[FZ J\"ulich]{Peter Gr\"unberg Institute (PGI-9), Forschungszentrum J\"ulich, 52425 J\"ulich, Germany, EU}
\author{Kenji Watanabe}
\affiliation[NIMS]{National Institute for Materials Science, 1-1 Namiki, Tsukuba, 305-0044, Japan}
\author{Takashi Taniguchi}
\affiliation[NIMS]{National Institute for Materials Science, 1-1 Namiki, Tsukuba, 305-0044, Japan}
\author{Gernot G\"untherodt}
\affiliation[RWTH Aachen University]{2nd Institute of Physics and JARA-FIT, RWTH Aachen University, 52074 Aachen, Germany, EU}
\author{Christoph Stampfer}
\affiliation[RWTH Aachen University]{2nd Institute of Physics and JARA-FIT, RWTH Aachen University, 52074 Aachen, Germany, EU}
\alsoaffiliation[FZ J\"ulich]{Peter Gr\"unberg Institute (PGI-9), Forschungszentrum J\"ulich, 52425 J\"ulich, Germany, EU}
\author{Bernd Beschoten}
\affiliation[RWTH Aachen University]{2nd Institute of Physics and JARA-FIT, RWTH Aachen University, 52074 Aachen, Germany, EU}
\email{bernd.beschoten@physik.rwth-aachen.de}
\phone{+49 (0)241 8027096}
\fax{+49 (0)241 8022306}

\title{Nanosecond spin lifetimes in single- and few-layer graphene-hBN heterostructures at room temperature}

\abbreviations{hBN}
\begin{document}

%%%%%%%%%%%%%%%%%%%%%%%%%%%%%%%%%%%%%%%%%%%%%%%%%%%%%%%%%%%%%%%%%%%%%
%% The "tocentry" environment can be used to create an entry for the
%% graphical table of contents. It is given here as some journals
%% require that it is printed as part of the abstract page. It will
%% be automatically moved as appropriate.
%%%%%%%%%%%%%%%%%%%%%%%%%%%%%%%%%%%%%%%%%%%%%%%%%%%%%%%%%%%%%%%%%%%%%

%%%%%%%%%%%%%%%%%%%%%%%%%%%%%%%%%%%%%%%%%%%%%%%%%%%%%%%%%%%%%%%%%%%%%
%% The abstract environment will automatically gobble the contents
%% if an abstract is not used by the target journal.
%%%%%%%%%%%%%%%%%%%%%%%%%%%%%%%%%%%%%%%%%%%%%%%%%%%%%%%%%%%%%%%%%%%%%
\begin{abstract}

We present a new fabrication method of graphene spin-valve devices which yields enhanced spin and charge transport properties by improving both the electrode-to-graphene and graphene-to-substrate interface. First, we prepare Co/MgO spin injection electrodes onto Si$^{++}$/SiO$_2$. Thereafter, we mechanically transfer a graphene-hBN heterostructure onto the prepatterned electrodes. We show that room temperature spin transport in single-, bi- and trilayer graphene devices exhibit nanosecond spin lifetimes with spin diffusion lengths reaching $\unit[10]{\mu m}$ combined with carrier mobilities exceeding $\unit[20,000]{cm^2/Vs}$.

\end{abstract}

%%%%%%%%%%%%%%%%%%%%%%%%%%%%%%%%%%%%%%%%%%%%%%%%%%%%%%%%%%%%%%%%%%%%%
%% Start the main part of the manuscript here.
%%%%%%%%%%%%%%%%%%%%%%%%%%%%%%%%%%%%%%%%%%%%%%%%%%%%%%%%%%%%%%%%%%%%%

In recent years, graphene has drawn strong attention because of measured spin-diffusion lengths of several microns at room temperature. Typical non-local spin-valve devices on Si/SiO${_2}$ substrates with charge carrier mobilities of several thousand $\unit{cm^2/Vs}$ exhibit spin lifetimes below \unit[1]{ns}~\cite{Tombros2007, Yang2011, Avsar2011, Volmer2013, Popinciuc2009, Jozsa2009, Maassen2011, Han2009, Han2010, Han2011}. In contrast, room temperature spin lifetimes above $\unit[1]{ns}$ have only been observed for epitaxial graphene on SiC\cite{Maassen2012} and for bilayer graphene devices (BLG) with low carrier mobility of $\unit[300]{cm^2/Vs}$ (Ref. \citenum{Yang2011}) or after post-processing of as-fabricated devices either by hydrogenation~\cite{Wojtaszek2013} or by oxygen treatment~\cite{volmer2014} which both, however, result in a decrease of the mobility. In graphene-based spintronics there is a device-oriented quest for combining long spin lifetimes with large carrier mobilities. While large carrier mobilities have been achieved in suspended structures~\cite{Guimaraes2012} and spin-valves on hexagonal boron nitride (hBN) crystals,~\cite{Zomer2012} the respective spin lifetimes only exhibit several $\unit[100]{ps}$. Recent studies indicate that the overall short spin lifetimes are most likely not a result of intrinsic spin scattering mechanisms in graphene but are rather caused and limited by the contact and interface properties of spin injection and detection electrodes.~\cite{Han2010,Maassen2012a,Volmer2013,Dlubak2012} This notion is supported by the observed increase of the spin lifetime with the contact-resistance-area products ($R_\text{c}A$) of both single-layer (SLG) and BLG devices.\cite{Volmer2013,volmer2014} In this context, it is important to note that the oxide barrier, which is needed for spin injection and detection, usually does not grow epitaxially on the graphene surface. MgO, for example, grows in a Volmer-Weber mode (island formation) if no additional wetting layer is used.~\cite{Wang2008} This island growth yields rather rough surfaces and additionally favors the formation of conducting pinholes between the overlaying ferromagnetic metal which is subsequently deposited and the underlaying graphene sheet. It has been suggested that these conducting pinhole states may be the bottleneck for spin transport when hybridizing with the graphene layer.~\cite{Volmer2013}

In this Letter, we present a new pathway for fabricating graphene spin-valves which diminishes some of the aforementioned shortcomings of the spin injection and detection contacts. In our approach, we first pattern MgO/Co electrodes which are deposited onto a silicon substrate. We emphasize that in contrast to all previous methods our MgO barrier is not deposited onto graphene but rather on top of the ferromagnetic Co layer. Thereafter, we mechanically press a graphene flake, which was previously transferred on hBN, onto the MgO surface of the electrodes. We show that this transfer technique allows for (1) nanosecond spin lifetimes with values up to $\unit[3.7]{ns}$ in trilayer graphene (TLG) devices which result from large ($R_\text{c}A$) values of the contacts and at the same time for (2) high carrier mobilities according to the enhanced charge transport properties by the mechanical contact of graphene to the overlaying flattening hBN layer.

\begin{figure}[tbp]
		\centering
		\includegraphics{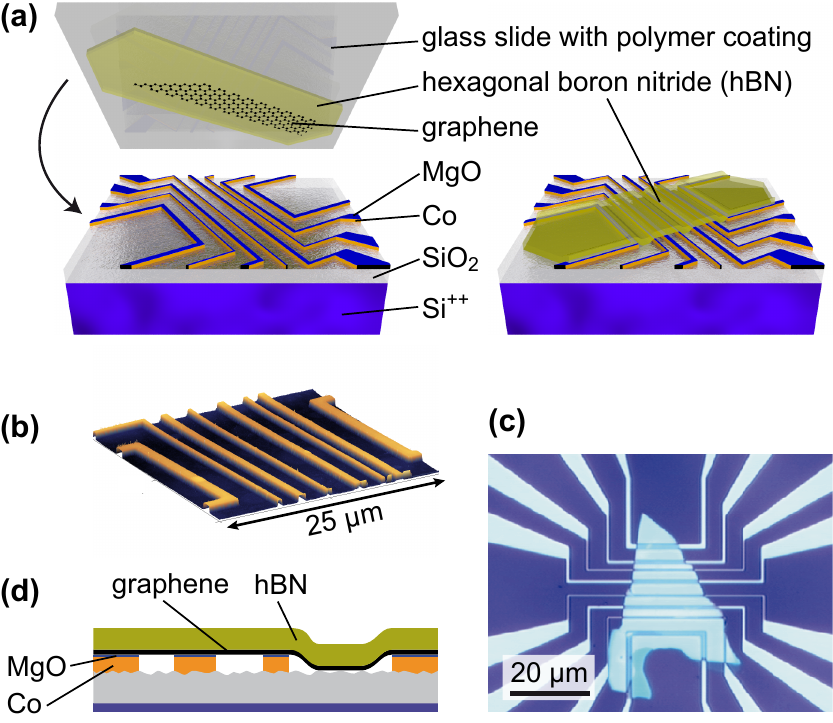}
	\caption{(a) Transfer technique for fabricating non-local graphene spin-valves (similar to Ref.~\citenum{Wang01112013}). (b) Scanning force microscope image of the prepared ferromagnetic electrodes. (c) Optical micrograph of a transferred graphene-hBN heterostructure onto MgO/Co electrodes. (d) Schematic cross section of the final device.}
	\label{fig:fig1}
\end{figure}

The new sample fabrication consists of two main steps (see Figures~\ref{fig:fig1}a and 1b) quite in contrats to the previously used top-down process. In a first step we prepare Co/MgO electrodes on a Si$^{++}$/SiO$_2$ substrate, where the Si$^{++}$ can be used as a back gate. The electrodes are defined by standard electron beam lithography and metallized by molecular beam epitaxy. We use $\unit[40]{nm}$ thick Co for spin injection and detection. Afterwards, we deposite $\unit[1]{nm}$ of MgO on top of the Co. This layer acts as an injection barrier to allow for efficient spin injection and detection. A scanning force microscopy (SFM) image of the prepared electrodes is shown in Figure~\ref{fig:fig1}b. The electrodes exhibit smooth surfaces and do not show fencing at their edges. Both is essential for a plane graphene to electrode interface.

The second step is illustrated in Figure~\ref{fig:fig1}a. We use exfoliated hBN which was transferred onto a glass slide covered by a polymer to pick up exfoliated graphene from a second Si$^{++}$/SiO$_2$ substrate (similar technique as described by \citeauthor{Wang01112013}~\cite{Wang01112013}). The thickness of the hBN varies between $\unit[60-80]{nm}$. Subsequently, the graphene-hBN heterostructure is mechanically transferred on top of the electrodes and the glass slide is thereafter removed by dissolving the polymer in acetone. A top-view optical image of the finished device is shown in Figure~\ref{fig:fig1}c.

\begin{figure*}[tbp]
		\centering
		\includegraphics{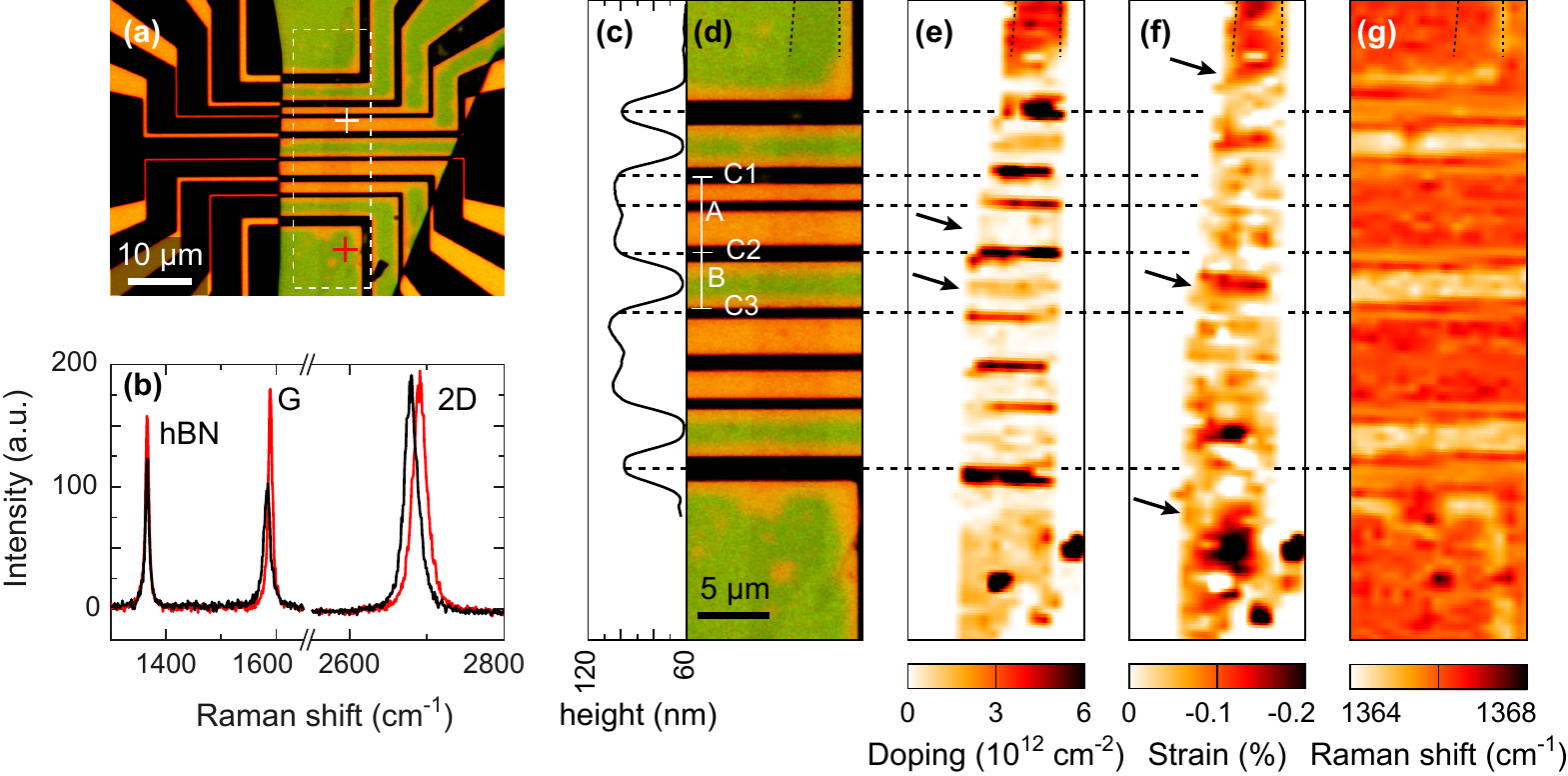}
	\caption{(a) Top view optical image of a Co/MgO/SLG-hBN device. (b) Raman spectra taken at the white and red crosses in panel (a). (c) SFM height profile scan showing the bending of the SLG-hBN. (d) Close up of white rectangular image detail in panel (a). (e) Doping and (f) strain distribution of the graphene flake. Values are extracted from the G and 2D-line positions of the Raman maps following Ref.~\citenum{Lee2012}. (g) Spatial image of the Raman hBN peak position (see panel (b)).}
	\label{fig:fig2}
\end{figure*}

As illustrated by the cross-section of our device (Figure~\ref{fig:fig1}d), the graphene-hBN heterostructure is suspended for small electrode spacings while it may bend down to the underlying SiO$_2$ surface (non-suspended) for larger spacings. The bending can easily be seen in the optical image in Figure~\ref{fig:fig2}a of a different SLG device where the optical contrast is encoded into a false-color scheme (orange for suspended and green for non-suspended), which even allows to visualize the underlaying graphene flake. An additional SFM line scan (Figure~\ref{fig:fig2}c) perpendicular to the electrodes in Figure~\ref{fig:fig2}d confirms this assignment.

To further analyze the quality of our transferred SLG-hBN heterostructure, we use micro-Raman spectroscopy to compare suspended with non-suspended regions. The respective spectra in Figure~\ref{fig:fig2}b have been taken at the positions of the white and red cross in Figure~\ref{fig:fig2}a. We observe three distinct peaks which can be attributed to hBN and the G- and 2D-line of graphene. For suspended graphene (red curve) we get peak positions (G-peak at $\unit[1583]{cm^{-1}}$ and 2D-peak at $\unit[2679]{cm^{-1}}$) which are very close to the values for pristine, freestanding graphene.~\cite{Lee2012} For the non-suspended graphene regions, however, we find a strong shift of both peak positions to larger wave numbers. This shift can be explained by local changes in doping and strain.~\cite{Lee2012}

We therefore recorded a Raman map over the device area shown in Figure~\ref{fig:fig2}d and use a vector decomposition of the G- and the 2D-peak shift \cite{Lee2012} to estimate doping and strain distributions in the graphene which are plotted in Figures~\ref{fig:fig2}e and 2f, respectively. We note that this method only allows to determine the carrier density but not its type.~\cite{Lee2012,Stampfer2007} For the suspended parts of the graphene flake there are only minor charge fluctuations ($< \unit[5 \times 10^{-11}]{cm^{-2}}$) visible while rather high doping is observed in all areas which are in direct contact with the electrodes. Moreover, doping occurs in the non-suspended regions which are in direct contact to the SiO$_2$ substrate. The doping may result from vacancies or charged defects in the MgO/Co electrodes and in the substrate exhibiting local electric fields which cause local doping.~\cite{Peres2010,DasSarma2011}

The local strain distribution is shown in Figure~\ref{fig:fig2}f. There is only small strain in suspended regions and in regions which are supported by the electrodes. As expected, we find larger strain in all non-suspended regions showing that the bending causes local strain. It is thus interesting to explore whether the local strain has any influence on the spin transport properties.~\cite{Huertas-Hernando2009} Furthermore, we note that the largest strain is measured in areas where graphene has direct contact to the SiO$_2$ surface (see arrows in Figure~\ref{fig:fig2}f). These findings are in agreement with previous results by \citeauthor{Lee2012} for graphene on SiO$_2$,~\cite{Lee2012} demonstrating that graphene exhibits less strain on hBN compared to SiO$_2$ substrates. The bending of the graphene-hBN heterostructure can also be seen as a peak shift of the hBN Raman line (Figure~\ref{fig:fig2}g) which can also be explained by local strain.~\cite{Gorbachev2011}

\begin{figure*}[tbp]
		\centering
		\includegraphics{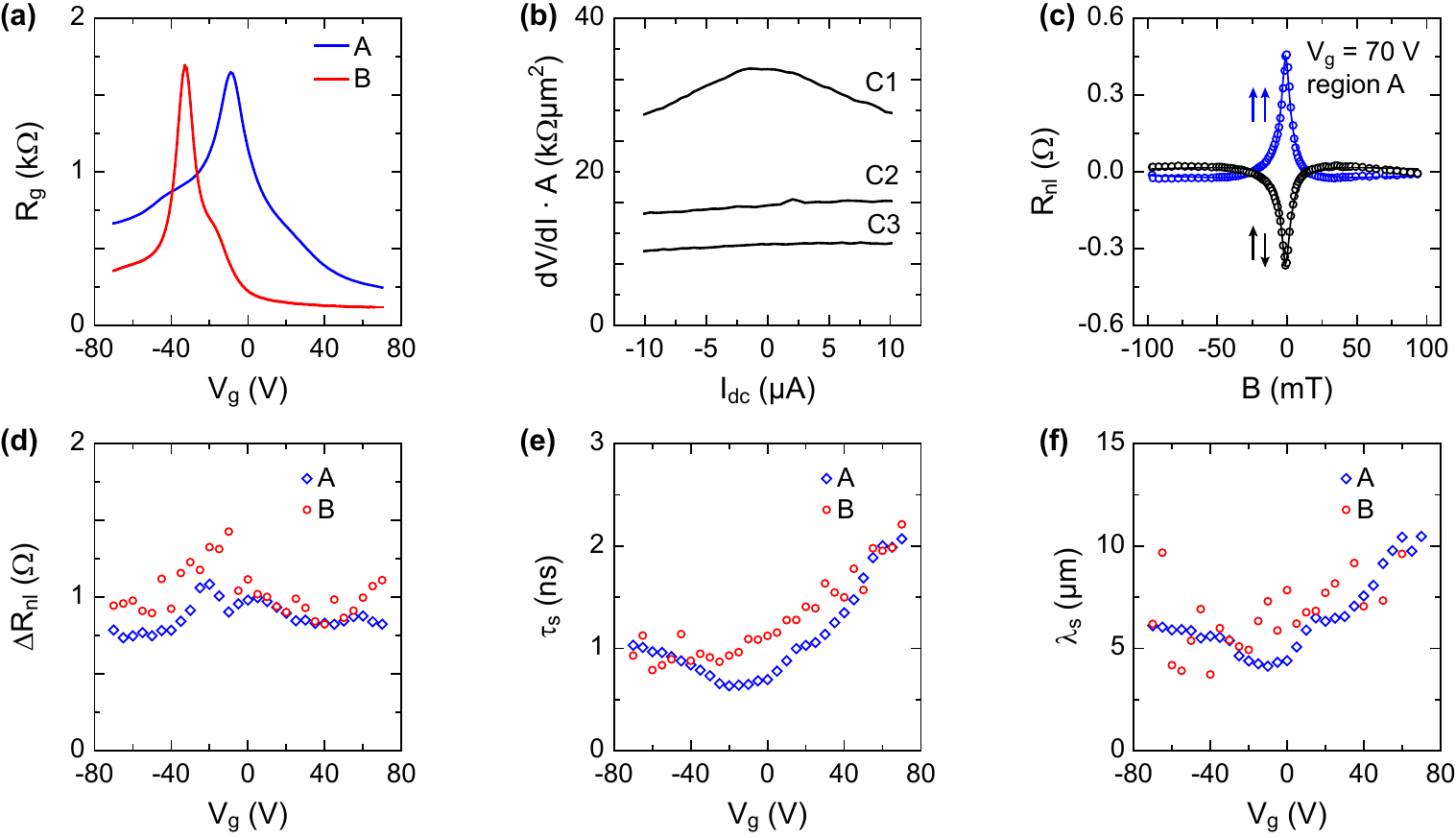}
	\caption{(a) Gate dependent graphene resistance of the SLG device from Figure~\ref{fig:fig2} for suspended region A (blue curve) and non-suspended region B (red curve). (b) $dV/dI\cdot A$ curves of contacts C1, C2 and C3 labeled in Figure~\ref{fig:fig2}d. (c) Hanle spin precession curves taken at $V_{\text{g}}= \unit[70]{V}$ in region~A. (d) Gate dependent spin signal for region~A (blue data points) and region~B (red data points). Respective gate dependent spin lifetimes are shown in panel (e) and spin diffusion lengths in panel (f).}
	\label{fig:fig3}
\end{figure*}

We next focus on spin and charge transport measurements on the SLG device presented in Figure~\ref{fig:fig2}. All transport measurements were performed at room temperature under vacuum conditions using a standard low frequency lock-in technique.\cite{note2} In Figure~\ref{fig:fig3}a we show the four-terminal back gate dependent resistance of the suspended region A (blue curve) and the non-suspended region B (red trace) (for assignment see Figure~\ref{fig:fig2}d). The charge neutrality point (CNP) of the suspended region is close to zero gate voltage while it is shifted to $V_g=\unit[-34]{V}$ for the non-suspended region. The shift of the latter results from a strong $n$-doping by the substrate whereas the suspended region A is only slightly doped which is both in accordance to the Raman analysis in Figure~\ref{fig:fig2}e.

We extract the carrier density by $n=\alpha (V_\text{g}-V_\text{g}^0)$ with $V_\text{g}$ being the back gate voltage and $V_\text{g}^0$ being the voltage to reach the CNP. The capacitive coupling constant $\alpha$ is different for suspended and non-suspended regions with respective values of $\alpha_\text{A} = \unit[3.5 \times 10^{10}]{V^{-1}cm^{-2}}$ and $\alpha_\text{B} = \unit[4.8 \times 10^{10}]{V^{-1}cm^{-2}}$.~\cite{note1}

The electron mobilities $\mu$ are determined from the gate dependent conductance $\sigma$ using $\mu = 1/e \cdot \Delta\sigma/\Delta n$. We obtain mobilities of $\unit[23,000]{cm^2/Vs}$ for the suspended region A and $\unit[20,000]{cm^2/Vs}$ for the non-suspended region B near the CNP. Compared to all previous room temperature carrier mobilities in graphene spin-valve devices on Si/SiO$_2$ substrates these values are more than a factor of 2 larger which highlights the high quality of our devices.~\cite{Yang2011}
We note that the given analysis is oversimplified as the graphene parts which are residing on top of the electrodes cannot be tuned by the gate voltage due to shielding of the back gate fields by the electrodes. The respective graphene resistance is thus a gate independent contribution to the total graphene resistance. This results in a smaller slope of $\sigma$ and thereby yield smaller mobilities when using the above conservative estimate.

We characterize all spin injection and detection electrodes (C1 to C3) of regions A and B (see Figure~\ref{fig:fig3}d) by their differential contact resistances ($dV/dI\cdot A$) with $A$ being the respective contact area.\cite{Volmer2013}. We observe large values for all contacts (Figure~\ref{fig:fig3}b). Only contact C1 exhibits the typical cusp-like dependence indicating tunneling behavior while the other contacts show a flat $dV/dI\cdot A$ curve which indicates transparent barriers. Interestingly, in previous studies transparent contacts could only be observed for contact resistance area products below $\unit[1]{k\Omega \mu m^2}$ for devices with a rather thick MgO layer ($\unit[2-3]{nm}$) which was directly deposited onto graphene.\cite{Volmer2013} We thus conclude that our thinner MgO barriers ($\unit[1]{nm}$) have better contact properties which might be explained by the weaker coupling to the graphene layer after the mechanical transfer of the graphene-hBN heterostructure onto the MgO surface.

Spin transport properties were measured in the standard four-terminal non-local Hanle geometry.~\cite{Jedema2002, Tombros2007, Lou2007} A typical measurement is shown in Figure~\ref{fig:fig3}c for region A at $V_\text{g} = \unit[70]{V}$ for parallel and antiparallel alignments of the Co magnetizations of the respective injector and detector electrodes after background subtraction. The spin signal $\Delta R_\text{nl}$ is given by the resistance difference at $B=0$~T. The Hanle depolarization curves are fitted by a simplified solution of the steady-state Bloch-Torrey equation~\cite{Torrey1956, Johnson1988, Fabian2007},

\begin{equation}
\label{eg:hanle}
\frac{\partial \vec{s}}{\partial t}\;=\;\vec{s}\times \vec{\omega}_0+D_{\text{s}}\nabla^2\vec{s}-\frac{\vec{s}}{\tau_{\text{s}}}\;=0,
\end{equation}

where $\vec{s}$ is the net spin vector, $D_{\text{s}}$ is the spin diffusion constant and $\tau_\text{s}$ is the spin lifetime. Additionally, $\vec{\omega}_0=g\mu_\text{B} \vec{B}/\hbar$ represents the Larmor frequency, where $\mu_\text{B}$ is the Bohr magneton, $\vec{B}=\vec{B}_\bot$ is the out-of-plane magnetic field and $g=2$ is the gyromagnetic factor. The spin diffusion length $\lambda_\text{s}$ is given by $\lambda_\text{s} = \sqrt{D_\text{s} \tau_\text{s}}$.

In Figure~\ref{fig:fig3}d we show $\Delta R_\text{nl}$ as a function of $V_g$ for the suspended region A (blue diamonds) and the non-suspended region B (red circles). For both regions the spin signal is on the order of $\unit[1]{\Omega}$ and shows a slight increase near the CNP. According to \citeauthor{Han2010} this increase is characteristic for tunneling contacts.~\cite{Han2010} As discussed above, two of the three electrodes show a flat $dV/dI$-curve (see Figure~\ref{fig:fig3}b), which indicates either transparent or intermediate contacts. We thus find that even intermediate contacts can exhibit the largest spin signal at the CNP.

In Figure~\ref{fig:fig3}e we plot the respective gate dependent spin lifetimes $\tau_\text{s}$. Most strikingly, we observe a strong increase of $\tau_\text{s}$ towards electron doping for $V_\text{g}>0$~V with maximum values exceeding 2~ns. The enhanced spin transport properties are also seen in the spin diffusion lengths which exceed $\unit[10]{\mu m}$ in both suspended and non-suspended regions (see Figure~\ref{fig:fig3}f). The latter values are larger than in all previous measurements on non-local spin valves including results on hBN ($\unit[4.5]{\mu m}$)~\cite{Zomer2012} and hydrogenated graphene ($\unit[7]{\mu m}$)~\cite{Wojtaszek2013}. The enhanced spin transport properties for increased doping values have also been observed in most previous experiments,~\cite{Jozsa2009,Avsar2011,Yang2011,Maassen2011,Han2011,Zomer2012,Han2012} but their origin is not completely understood.~\cite{Zhang2011,Zhang2012,Diez2012}. Surprisingly, there is no distinct difference in $\tau_\text{s}$ and $\lambda_\text{s}$ between the suspended and the non-suspended region indicating that even in our new devices the spin lifetime is still limited by contact-induced spin dephasing which hinders to explore relevant spin dephasing and spin relaxation mechanisms in the graphene layer itself.

\begin{figure}[tbp]
		\centering
		\includegraphics{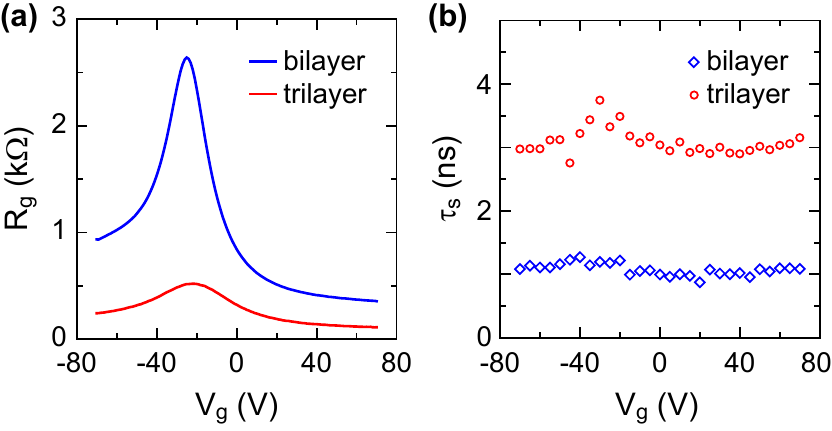}
	\caption{(a) Graphene resistance as a function of gate voltage for non-suspended BLG (blue curve) and TLG (red curve) devices. (b) Room temperature spin lifetime vs back gate voltage for BLG (blue diamonds) and TLG device (red circles).}
	\label{fig:fig4}
\end{figure}

It is interesting to compare these SLG results to few-layer graphene devices. By the same fabrication method we therefore additionally prepared BLG and TLG devices and show results on charge transport and spin lifetimes in Figures~\ref{fig:fig4}a and 4b, respectively. The data were taken on non-suspended regions which again result in a shift of the CNP towards negative gate voltages (see Figure~\ref{fig:fig4}a). We estimate electron carrier mobilities of $\unit[9,000]{cm^2/Vs}$ for BLG and $\unit[10,000]{cm^2/Vs}$ for TLG where we used $\alpha = \unit[4.9 \times 10^{10}]{V^{-1}cm^{-2}}$ for both. As discussed above, these values should be taken as a lower limit estimate. In contrast to the SLG device, we observe a completely different density dependence of the respective spin lifetimes (Figure~\ref{fig:fig4}b). While $\tau_\text{s}$ depends only weakly on gate voltage it becomes largest at the CNP reaching $\unit[3.7]{ns}$ for the TLG device, which is the largest room temperature value in graphene-based non-local spin transport to date. A similar gate voltage dependence was previously also observed in other graphene spin-valve devices~\cite{Maassen2011,Han2011,Zomer2012,Wojtaszek2013} including SLG. We therefore do not attribute this behavior to the number of graphene layers.

\begin{figure}[tbp]
		\centering
		\includegraphics{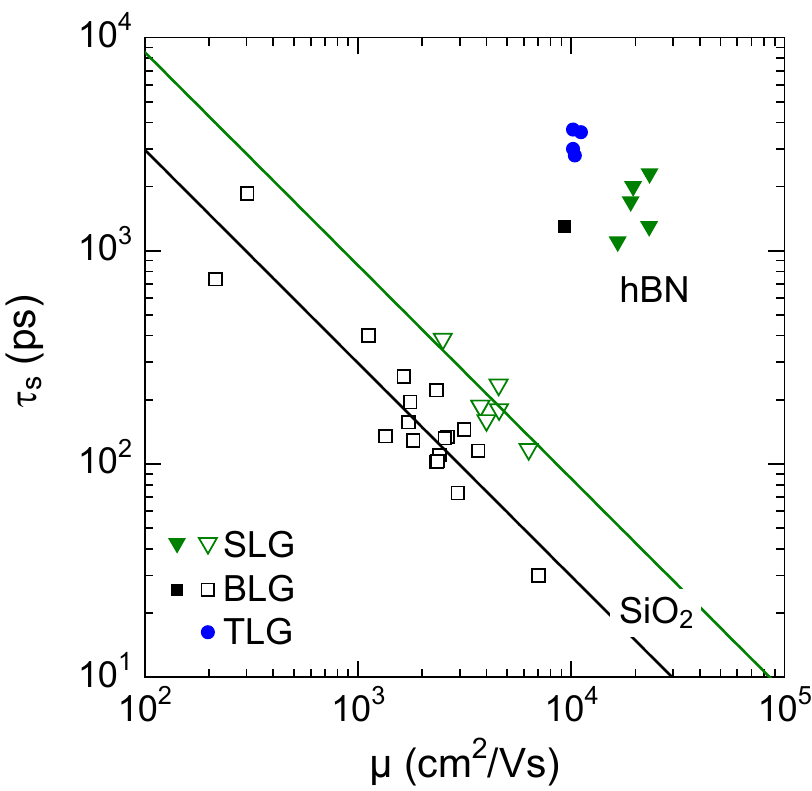}
	\caption{Room temperature spin lifetime as a function on carrier mobility obtained on single- and few-layer graphene non-local spin-valve devices for Co/MgO spin injection and detection electrodes. Results from the present study are depicted by the filled symbols while previous results on single and bilayer graphene (open symbols) are taken from Ref. \citenum{Yang2011} and \citenum{Volmer2013}.}
	\label{fig:fig5}
\end{figure}

Figure~\ref{fig:fig5} summarizes our results. It shows the dependence of $\tau_\text{s}$ on the electron mobility on a log-log scale. Data points from the present study are shown by filled symbols. In total we measured 10 regions of 5 devices. In 9 regions we obtain maximum spin lifetimes above $\unit[2]{ns}$ and mobilities above $\unit[10,000]{cm^2/Vs}$ which demonstrates the reproducibility of the device performance by our fabrication method. For easier comparison we include results on SLG and BLG which some of us had previously measured.\cite{Yang2011,Volmer2013} All latter devices were  prepared on Si/SiO$_2$ by a conventional top-down fabrication method in which the MgO barriers are directly evaporated onto graphene. The overall improvements of the performances of our new devices are striking. These new devices exhibit spin lifetimes which are two orders of magnitude longer than the previous BLG devices (squares) with the largest mobility of $\unit[8,000]{cm^2/Vs}$ that only yielded $\tau_\text{s}= \unit[30]{ps}$.\cite{Yang2011} While a nanosecond spin lifetime was obtained in BLG devices with mobilities as low as $\unit[300]{cm^2/Vs}$ we now obtain nanosecond spin lifetimes with mobilities which are almost two order of magnitude larger. We clearly attribute the increase in the mobilities to the hBN substrate while we relate the increase of the spin lifetimes to improved contact (i.e. electrode interface) properties according to our advanced transfer technique onto prepatterned electrodes which has several advantages over previous methods. Firstly, the contact region of the graphene is never exposed to an electron beam, which most likely lead to a smaller number of spin-scattering centers in graphene.~\cite{Childres2010} Secondly, the interface between graphene and MgO is expected to be of higher quality, since more aggressive cleaning procedures can be used for removing resist residues from the lithography step. Furthermore, there is an island growth of MgO on graphene~\cite{Wang2008} whereas it can grow fully epitaxial on Co.~\cite{Yuasa2006} Although our present MgO layers do not grow epitaxially they exhibit more homogeneous barriers as in previous studies. The homogeneity of the MgO layer does not only lead to an improved spin injection efficiency but also prohibits direct contact of Co atoms to graphene which is known to induce 3d-like hybridized states in graphene and can yield strong spin scattering.~\cite{Varykhalov2012}

In summary, we presented a new way of fabricating graphene-hBN spin-valve devices where we mechanically transfer the graphene onto predefined Co/MgO electrodes. All single-layer, bilayer and trilayer graphene devices exhibit nanosecond spin lifetimes up to 3.7~ns with carrier mobilities exceeding $\unit[20,000]{cm^2/Vs}$ and spin diffusion lengths above $\unit[10]{\mu m}$. Our presented transfer method can be applied to epitaxial oxide barriers, which are expected to yield even longer spin lifetimes. This ultimately paves the way to explore intrinsic spin transport properties and to realize promising devices in highest mobility graphene.

\begin{acknowledgement}
We thank S. G\"obbels for helpful discussions and S. Engels for help with the transfer process.
The research leading to these results has received funding from the DFG through FOR-912,
the People Programme (Marie Curie Actions) of the European Union's Seventh Framework Programme FP7/2007-2013/ under REA grant agreement n°607904-13 and the Graphene Flagship (contract no. NECT-ICT-604391).
\end{acknowledgement}

%%%%%%%%%%%%%%%%%%%%%%%%%%%%%%%%%%%%%%%%%%%%%%%%%%%%%%%%%%%%%%%%%%%%%
%% The same is true for Supporting Information, which should use the
%% suppinfo environment.
%%%%%%%%%%%%%%%%%%%%%%%%%%%%%%%%%%%%%%%%%%%%%%%%%%%%%%%%%%%%%%%%%%%%%
%\begin{suppinfo}

%\end{suppinfo}

%%%%%%%%%%%%%%%%%%%%%%%%%%%%%%%%%%%%%%%%%%%%%%%%%%%%%%%%%%%%%%%%%%%%%
%% The appropriate \bibliography command should be placed here.
%% Notice that the class file automatically sets \bibliographystyle
%% and also names the section correctly.
%%%%%%%%%%%%%%%%%%%%%%%%%%%%%%%%%%%%%%%%%%%%%%%%%%%%%%%%%%%%%%%%%%%%%
%\bibliography{bibliography}

\begin{mcitethebibliography}{40}
\bibitem[not()]{note1}
In contrast to graphene which was directly deposited onto Si/SiO$_2$ we have to
  consider an additional dielectric (air) between the graphene and the surface
  of the substrate as well as the partial shielding of the electric field by
  the nearby electrodes. We therefore simulate the capacitive coupling of our
  devices using \textsc{Comsol} for both suspended and non-suspended regions
  with different electrode spacings which gives an estimate for $\alpha$\relax
\mciteBstWouldAddEndPuncttrue
\mciteSetBstMidEndSepPunct{\mcitedefaultmidpunct}
{\mcitedefaultendpunct}{\mcitedefaultseppunct}\relax
\EndOfBibitem
\end{mcitethebibliography}

\providecommand{\latin}[1]{#1}
\providecommand*\mcitethebibliography{\thebibliography}
\csname @ifundefined\endcsname{endmcitethebibliography}
  {\let\endmcitethebibliography\endthebibliography}{}

\end{document}